\documentclass{nature}

\usepackage{graphicx}
\usepackage{booktabs}
\usepackage[flushleft]{threeparttable}
\usepackage{amsmath,amssymb}
\def\be{\begin{equation}}
\def\ee{\end{equation}}
\newcommand{\ba}{\begin{eqnarray}}
\newcommand{\ea}{\end{eqnarray}}
\usepackage{siunitx}
\title{Formation of surface nanodroplets under controlled flow conditions}

\author{Xuehua Zhang$^{1,2}$, Ziyang Lu$^{1}$, Huanshu Tan$^{2}$, Lei Bao$^{1}$, Yinghe He$^{3}$, Chao Sun$^{2}$, Detlef Lohse$^{2}$}

\begin{document}

\maketitle

\begin{affiliations}
 \item School of Civil, Environmental and Chemical Engineering, RMIT University, Melbourne, VIC 3001, Australia,
 \item Physics of Fluids Group, MESA+ Institute, and J. M. Burgers Centre for Fluid Dynamics, University of Twente, 7500 AE Enschede, The Netherlands,
 \item Centre for Biodiscovery and Molecular Development of Therapeutics, College of Science, Technology  Engineering, James Cook University, Townsville City QLD 4811, Australia,
\end{affiliations}

\begin{abstract}
Nanodroplets on a solid surface (i.e. surface nanodroplets) have practical implications for high-throughput chemical and biological analysis, lubrications, lab-on-chip devices, and near-field imaging techniques.  Oil nanodroplets can be produced on a solid-liquid interface in a simple step of solvent exchange
in which  a good solvent of oil is displaced by a poor solvent.   In this work, we  experimentally and theoretically investigate the formation of nanodroplets by the solvent exchange process under well-controlled flow conditions.  
We find that the contact angle of the nanodroplets is
 independent of the flow condition. 
However, there are significant effects from the flow rate and the flow geometry on the droplet size. 
We develop a theoretical framework to account for these effects. The main idea is that the 
droplet nuclei are exposed to an oil oversaturation pulse during the exchange process. 
The analysis gives 
 that the volume of the nanodroplets increases with the Peclet number $Pe$ of the flow
 as $\propto Pe^{3/4}$,
 which is in good agreement with our experimental results. In addition, 
 at fixed flow rate and thus fixed Peclet number, 
 larger and less homogeneously distributed droplets  formed at less narrow  channels, 
  due to convection effects originating
  from the density difference between the two solutions of the solvent exchange.  The understanding from this work provides valuable guidelines for producing surface nanodroplets with desired 
   sizes by controlling the flow conditions.  
\end{abstract}

Nanoscale droplets on a substrate are an
essential element for a wide range of applications,
namely  lab-on-chip devices, simple and highly-efficient miniaturised reactors for concentrating products, high-throughput single-bacteria or single-biomolecular analysis,  and high-resolution imaging techniques, amongst others \cite{antonio2009,chiu2009,shemesh2014,meckenstock2014}. Quite
some  effort has been devoted to produce a large amount of nanodroplets in a  controlled way. The current techniques include trapping by microcavities, emulsion direct adsorption, microprinting and others \cite{day2012}. The solvent exchange process is a simple and generic approach for producing droplets or bubbles at solid-liquid interfaces that are only several tens to hundreds nanometers in height, or a few femtoliter in volume \cite{lou2000,zhang2007nanodroplet,zhang2012softmatter,belova2013,craig2011softmatter}. It has 
 attractive advantages,  such as its capability of producing  a large number of nanodroplets in one simple step, and its generality in chemical composition of the droplet liquid, and flexibility in aspect ratio of the droplets and spatial structure or size of the substrate \cite{zhang2012softmatter,yang2014}.   

For the formation of surface nanodroplets by solvent exchange, a hydrophobic substrate is exposed sequentially to two miscible solutions of oil, where the second solvent has a lower solubility of oil than the first.  Such solubility difference leads to  supersaturation of 
the liquid with oil during the solvent exchange and consequently to the nucleation of 
 nanodroplets on the substrate.   The analogue technique in a bulk system is called {\it solvent shifting} or {\it nanoprecipitaion} through the Ouzo effect \cite{vitale2003}, which has been increasingly applied to obtain nanodroplets in a surfactant-free emulsion \cite{ma2015}, monodispersed polymeric nanoparticles with precisely-controlled sizes \cite{aubry2009,lepeltier2014,yang2014,stepanyan2012}, or assemble colloidal particles on a microscale \cite{grau2015}.

 Although the chemical composition 
 of the solutions has been used to adjust the average size of the droplets \cite{yang2014}, the flow properties dramatically complicate the formation of nano\-droplets by the solvent exchange. 
The reason is that the mixing of the two liquids strongly depends on the flow conditions \cite{stroock2002,al-Housseiny2012}. 
 Inspired by the pattern formation of mineral aggregates from liquid displacement under a quasi-2D flow conditions \cite{haudin2014,steinbock2014}, we may be able to control the droplet nucleation and growth by the flow conditions in a well-defined flow system. In this work, we theoretically and experimentally investigate 
 the effects of the 
 flow conditions on the formation of surface nanodroplets.    We find
  that 
  the averaged volume of surface nanodroplets increases with the Peclet number 
  as $\propto Pe^{3/4}$, in good agreement with the experiments.
    As far as we know, this work is the first attempt to quantitatively understand
    the effects of the flow conditions during the solvent exchange on the formation of surface nanodroplets. 
    
\section*{Results and discussion}

\subsection{Droplet volume dependence  on the  flow rate}

The geometry of the solvent exchange channel and process is shown in schematic drawings in Figure~\ref{afm}(a).   
 Three fluid channels with different heights were used and their dimensions are listed in Table~\ref{tab:geometry}. 
 During the solvent exchange process, solution A
 (50\% ethanol aqueous solution saturated with polymerisable oil,  with high oil solubility) 
 was displaced by solution B  (oil-saturated water, with low oil solubility). The injection of solution B was performed at a constant flow rate $Q$ controlled by a syringe pump.  Once the solvent exchange was completed, the nanodroplets on the substrate were cured by photopolymerization. 

Figure~\ref{afm}(b) show
a representative AFM image of the polymerised nanodroplets. 
The polymerised droplets are spherical caps with a certain size distribution. The plot in Figure~\ref{afm}(c) shows that the droplet height increases monotonically from 10 nm to 300 nm as the lateral size increases from 2 $\mu$m up to 10 $\mu$m, in consistency with the previous reports \cite{zhang2012softmatter}.  The contact angles of those polymerised droplets lie mainly between the
macroscopic receding ($8^o$) and advancing ($19^o$) contact angles of the system\footnote{Strictly speaking, the contact angle of a liquid droplet before the polymerisation is about $1-2^o$ larger than that of its polymerised counterpart due to curing shrinkage. }. Clearly,
 the flow rate does not influence the contact angle of the surface nanodroplets. % neither does it influence the droplet number density.  

The droplet volume is calculated 
 from its lateral diameter in the optical images of the polymerised droplets in Figure~\ref{optical}, and the corresponding contact angle in Figure~\ref{afm}. Figure~\ref{optical}(d) shows the probability distribution function (PDF) of the droplet volume at different flow rates in the narrowest channel, 
where the distribution of the droplet volume became wider at a faster flow rate.  To further examine effects of the flow rate on the droplet size,  we measured the lateral diameter of droplets that were produced from different flow rates at all three channels.   The averaged lateral diameter of the droplets versus the flow rate is plotted in Figure~\ref{optical}(e), which shows a fast increase of the droplet size with an increase in the flow rate for all three channel heights. 
For instance, in the narrow channel the spatially averaged droplet diameter increased from 2 $\mu$m to 10 $\mu$m (20 nm to 300 nm in height) as the flow rate increased from 100 $\mu$l/min to 2400 $\mu$l/min. It shows the same trend for the two less narrow channels,
but the absolute values are larger on average.

 %We obtain values between 0.1 to 15 femtoliter.  
  We analysed all  droplet volumes over a surface area of 0.35 mm$^2$  and obtained the averaged droplet 
  volume per unit area of $\mu$m$^2$.
  The plot in Figure~\ref{optical}(f) shows a sharp increase of the averaged droplet volume with
  increasing  flow rate. The same data are shown in  
   Figure~\ref{optical}(g) (in a log-log plot)
  versus the Peclet  number
   \be 
Pe = {\bar U h\over D}  = {Q \over w D} 
\label{pe}
\ee 
 of the flow, where $D$ is the diffusion constant.  
 The data can be described with the scaling law $\propto Pe^{3/4}$. Later we will show that this 
 scaling law 
  between the droplet volume and the $Pe$ number is in a good agreement with the theoretical prediction. 

It is crucial to identify the formation mechanism of the droplets before we develop a theoretical model to understand the effect of the flow rate.  In our tertiary system of ethanol-oil-water, surface nanodroplets may nucleate on the surface, or through the standard ``ouzo effect'', namely
 nucleate in the bulk liquid 
 \cite{vitale2003, zhang2008droplets, schubert2011, yan2014, stepanyan2012}, 
 and only later
  adsorb onto the surface.  We analysed the droplet volume and the surrounding area of droplet-depleted zone, following the modified Vorono\"{i} tessellation method in our previous work \cite{lhuissier2014}. 
  %Fig.\ref{voronoi} shows 
  In Supporting Figure 1, we show the relation between the depleted area 
  and the area of the corresponding footprint of nanodroplets produced at
   different flow rates. For the relatively large nanodroplets, 
   the depleted area is proportional to the droplet footprint area. This correlation clearly suggests that surface nanodroplets are not from random adsorption of emulsion droplets, but from heterogenous nucleation and subsequent diffusion-driven growth. In this process, the oil dissolved in the bulk is consumed by the growth of the droplets, leading to the droplet-depleted area in the surrounding region.  The same diffusion-driven mechanism and the same correlation between the depleted area and the footprint area
   was also observed (and derived) for the spatial arrangement of surface nanobubbles formed by the solvent exchange method \cite{lhuissier2014}.  

\section*{A simple theoretical model}
\subsection{Mathematical description of the solvent exchange process}

As shown in the schematic drawings of Figure~\ref{afm}, the flow cell consists of a channel with height $h$
and  channel width $w$. The maximal flow velocity is $U$, and mean flow velocity $\bar U$. The resulting flow rate is $Q = hw \bar U$, its Reynolds number $Re=\bar U h/ {\nu} = Q/ (w\nu ) $. For the theory we will focus on
 laminar flow, i.e. $Re \lesssim 1$, which, as seen from Table~\ref{tab:geometry},  is justified for all channels.
 Furthermore,
   we will neglect the  density contrast between the ethanol solution and the water (factor about 0.9), 
   which, as we will see below, strictly speaking is only justified  for the narrow channel case $h=0.33$ mm.

In principle, the solvent exchange process is described by the advection diffusion equations for the ethanol solution and water and the oil 
dissolved in it, with the no-slip boundary conditions on all walls.
The initial conditions are such that in the left part of the channel there is no ethanol dissolved in the water,
but oil up to its saturation concentration $c_{s,wat}$ in water. 
In the right part of the channel we have ethanol solution with dissolved
oil at saturation concentration $c_{s,eth} > c_{s,wat}$. At time $t=0$ the 
interface  between the two parts of the flow is 
assumed to be sharp. 
From $t=0$ on,  the flow is driven by a  pump such that
the flow rate $Q$ is constant. 
The interface will then develop a  parabolic shape, corresponding to the laminar flow situation. 
It will smoothen out with advancing time, see the sketch Figure~\ref{profile}(a).
Note that downstream in the region which
was initially filled with ethanol solution the front will hit the surface nearly parallel to the surface, due to the no-slip boundary
condition. The width of the front is given by the diffusion process of oil (and ethanol) towards the water and water
towards the ethanol. 
The boundary conditions for the oil are no flux boundary conditions at the top and bottom wall.
Once there is oil droplet nucleation, the oil concentration equals $c_{s,wat}$ at the droplet-water interface.

For nucleating and growing  oil droplets the most relevant quantity is the oil oversaturation 
\be
\zeta(t) = {c_\infty (t) \over c_s} -1 . 
\ee
As the liquid is saturated in the ethanol phase, there we have $c_\infty = c_{s, eth}$ and thus $\zeta = 0$. 
The same holds in the water oil phase, $c_\infty = c_{s,wat}$ and thus 
also $\zeta = 0$. However, in the broadening front
around the interface we have oversaturation 
$\zeta >0$, as oil diffuses from the ethanol-rich phase towards the water-rich phase, in which it is less soluble. The maximum
oversaturation is 
\be
\zeta_{max}  = {c_{s,eth}  \over c_{s,wat}} -1  > 0 .  
\label{zetamax}
\ee

At fixed position downstream (what initially is the ethanol phase) we first have no oversaturation, $\zeta =0$. 
Then the front is passing by during which $\zeta (t) > 0$ so that oil droplets can nucleate and grow.
In the end the ethanol is basically fully replaced by the oil-saturated water and then again $\zeta = 0$. 
This front is characterised by the maximum $\zeta_{max}$ and by some temporal width $\tau$, see Figure~\ref{profile}(b). 
One may be tempted to argue that this temporal width $\tau$ depends on the flow velocity. However, note that 
independent of the flow velocity the nucleating droplets on the surface are horizontally hit by the mixing front, given a no-slip boundary condition. 
We therefore argue that in our laminar flow cell $\tau$ is independent of $U$ and purely given by the diffusion
process, suggesting $\tau \sim { h^2 \over D}$. For $h = 0.33$ mm as in the narrow channel experiments and $D = 1.6 \times 10^{-9} m^2/s$ 
we get $\tau \approx 60 s$.

\subsection{Growth of a nucleated droplet}
Once the oversaturation front passes, droplets nucleate and grow.  
Here we focus on an individual droplet. We assume that there is no pinning and that the contact angle $\theta$ is thus constant.
The size of the droplet is characterised by its lateral extension $L$. Alternatively, we can use the radius of curvature $R$ of 
the oil-water interface as a characterisation of the droplet size, which we will do here. Then 
$
Vol \sim R^3
$
for the droplet volume, as $\theta$ is constant. Assuming different growth modes 
as stick-slip or stick-jump \cite{zhang2015} 
would only 
change prefactors, but not the essence of below derivation.

Strictly speaking, for the droplet growth no symmetry holds: The axial symmetry, which is obeyed for a drop diffusively growing in  still liquid, is broken by the flow direction. Nonetheless, to obtain the scaling relations we can still assume 
the  diffusive growth equation even in spherical symmetry,
\be
\dot m = 4\pi \rho_{oil} R^2 \dot R = 4\pi D R^2 \partial_r c|_R. 
\label{mdot}
\ee
In this laminar flow situation the concentration gradient at the interface $\partial_r c|_R$
is given by the oil concentration difference
between the oil concentration in the flow $c_\infty$ and at the interface $c_{s,wat}$ and by the thickness $\lambda$
of the concentration boundary layer, for which we assume a Prandtl-Blasius Pohlhausen's type behavior 
\cite{sch00,gro04} 
as appropriate for laminar flow. 
Then $\lambda \sim {R \over \sqrt{Pe}}$.
I.e., we have 
\be
\partial_r c|_R \sim {c_\infty (t)  - c_{s,wat} \over \lambda } \sim c_{s,wat} {\zeta (t) \over \lambda } \sim 
c_{s,wat} \sqrt{Pe} ~ R^{-1} \zeta (t)    . 
\label{grad}
\ee
Plugging this into equation \eqref{mdot} we obtain a simple ODE for $R(t)$, namely
\be
R \dot R \sim {D c_{s,wat} \over \rho_{oil} } \sqrt{Pe} ~ \zeta (t) ,  
\label{ode}
\ee
which can easily be integrated respectively from 0 to the final radius $R_f$ or from 0 to $t=\infty $, giving
\be
R_f  \sim \left( 
{D c_{s,wat} \over \rho_{oil} } \zeta_{max} \tau Pe^{1/2}  
\right)^{1/2} . 
\ee
Here we have used $\int_0^\infty  \zeta (t) dt = \zeta_{max} \tau $.  
Using our prior assumptions $\tau \sim {h^2 \over D}$ (in particular that it is flow rate independent) and equation \eqref{zetamax} on $\zeta_{max}  $
we obtain 
\be
Vol_f \sim R_f^3 \sim h^3 \left (  { c_{s,wat} \over \rho_{oil}   }\right)^{3/2}   \left( {c_{s,eth}  \over c_{s,wat}} -1 \right)^{3/2} ~ Pe^{3/4}
\label{vol_f}
\ee
for the final volume of the droplet after the solvent exchange. 
The scaling $Vol_f \sim Pe^{3/4}$  of our theoretical model
 is in good agreement with the experimental data shown in Figure~\ref{optical}(g), given that the 
 droplet number density is flow rate independent.

 \section*{Buoyance driven convection effects for the less narrow channels}
 
 We now examine the effect of the channel height on the droplet formation.  First of all we characterised the flow in the channels by using fluorescent microscopy, while water was dyed green to assist the visualisation. Top-view videos of the entire exchange process are provided in the Supporting Videos.  The snapshots in Figure~\ref{snapshot}(a) show that in the narrowest channel the water displaced the ethanol solution in a  smooth and continuous manner. The fluorescent intensity on a specific location increased with time smoothly, as the concentration of water increased in the liquid phase.  The top-view snapshots in Figure~\ref{snapshot}(b) and Figure~\ref{snapshot}(c) show the flow patterns for the two less narrow  channels at the same mean flow velocity of $\bar U$ of 0.36 mm/s.  Straight and regular fingers were clearly visible in the channel with $h=0.68$ mm, while the flow already developed whirling patterns for $h= 2.21$ mm. 
The time evolution of the fluorescent intensity of the dye in water shows some deflections and jumps in the intensity,  indicating the abrupt change in water contents due to the non-uniform mixing.  Such flow features are in contrast to the smooth flow in the narrow channel at $h=0.33$ mm.
Note that in all three cases the flow is still laminar, see Table~\ref{tab:geometry}.

 The reason for the different flow patterns in the less narrow  channels is that for  them 
we must  consider the density 
 difference between the  two miscible liquids \cite{lajeunesse1997,bischofberger2014}. The density of water is 1 g/ml  while the density of the ethanol aqueous solution is $\sim$0.90 g/ml.
 At solvent exchange, at the bottom side of the channel above the plate, the lighter ethanol will be pushed
 below the entraining heavier water, potentially  leading to some buoyancy driven 
 convection rolls.
 To estimate when these convection rolls set in, we can define a ``Rayleigh number''
\begin{equation}
Ra = \frac{\Delta \rho g (h/2)^3}{\mu D_{e,w}},
\label{Ra}
\end{equation}
where the density difference $\Delta \rho$ is $0.1$g/ml,  the 
 gravitational acceleration $g$ is 9.8m/s$^{2}$, $\mu$ is  the dynamic viscosity of ethanol solution,
  and the mass diffusion coefficient of ethanol and water $D_{e,w}$ at 300K is $1.6\times10^{-9}$m$^2$/s. 
 Convection  only occurs at the lower half of the channel where heavy liquid (water) is above light liquid (ethanol), see Figure~\ref{convection}(a). Therefore we take $h/2$ as vertical length-scale in  equation \eqref{Ra}. The resulting estimated Rayleigh numbers are $\approx 1.1 \times 10^3$, $1.0 \times 10^4$, and $3.5 \times 10^5$ for the channel heights of $0.33$ mm, $0.68$ mm and $2.21$ mm, respectively. 
So the convection rolls (in the top view seen as stripes) only occur in the two less narrow 
 channels, where the Rayleigh number is larger than the critical Rayleigh number 1708 \cite{faber1995fluid}.
The existence of convection rolls for the two less narrow channels
also explain why for fixed flow rate and thus fixed  Peclet number,  
 the average droplet size depends on the channel height, 
 in particular for high flow rates, as seen in Figure~\ref{optical}(a)(b)(c).
 The convection rolls lead to a better mixing between ethanol and water,
 and thus better transport of oil towards the substrate  and consequently to larger oil droplets.

The development of the secondary flow  at a high flow rate may lead to some patterns of the droplets. Indeed, 
 we observed that some of the droplets produced in the least narrow 
  channel line up along the flow direction as shown in  Figure~\ref{convection}(b). Such lines of droplets also formed at 
  larger
   flow rate in the channel with $h=0.68$ mm. They form when the convection rolls hit the surface, bringing down oil rich liquid. All these features do not develop when we turn the channel by $90^0$ (see Supporting Figure 2), through which we eliminate buoyancy effects.

\section*{Conclusions}

In summary, we theoretically and experimentally investigate the formation of surface nanodroplets by solvent exchange under well-controlled flow conditions.  We found that although the contact angle of surface nanodroplets is independent of flow conditions, the flow rate and flow geometry have significant effects on the droplet size. We developed a theoretical framework for the solvent exchange process, whose 
result  is in good agreement with the experimental results, 
namely  that the droplet volume increases with $Pe^{3/4}$.  
Increasing the channel height (for given flow rate and thus given $Pe$) 
can induce convection driven by the density difference between water and ethanol, 
leading to larger droplets and an inhomogeneous droplet nucleation pattern, which 
reflects the convection rolls. 
The results presented in this work provides valuable guideline for the device design to generate surface nanodroplets with some desired  sizes.

\section*{Experimental section} 
\subsection{Substrate and solutions}

A stock solution containing monomer and initiator was prepared by mixing 1, 6-hexanediol di-acrylate (HDODA) (80 \%, Sigma-Aldrich) with 2-hydroxy-2-methylpropiophenone ( 97 \%, Sigma-Aldrich) in the ratio of 10:1. This solution of monomer precursors served as oil phase. 4 ml of the monomer solution was added into 100 ml ethanol/water (50 v\%: 50 v\%) solution, and the bottom phase of the liquid was the solution A.  Solution B was water saturated with HDODA.  The hydrophobic substrate of OTS-Si was prepared and cleaned by following the protocol reported in previous work \cite{zhang2012softmatter}. Before use,  the OTS-Si was cleaned with chloroform, sonicated in ethanol, and dried with nitrogen.  The advancing and receding contact angles of water were $112^o$ and $98^o$, respectively.

\subsection{Preparation, polymerisation and characterisation of nanodroplets}

\begin{figure*}
\centerline{\includegraphics[width=.9\textwidth]{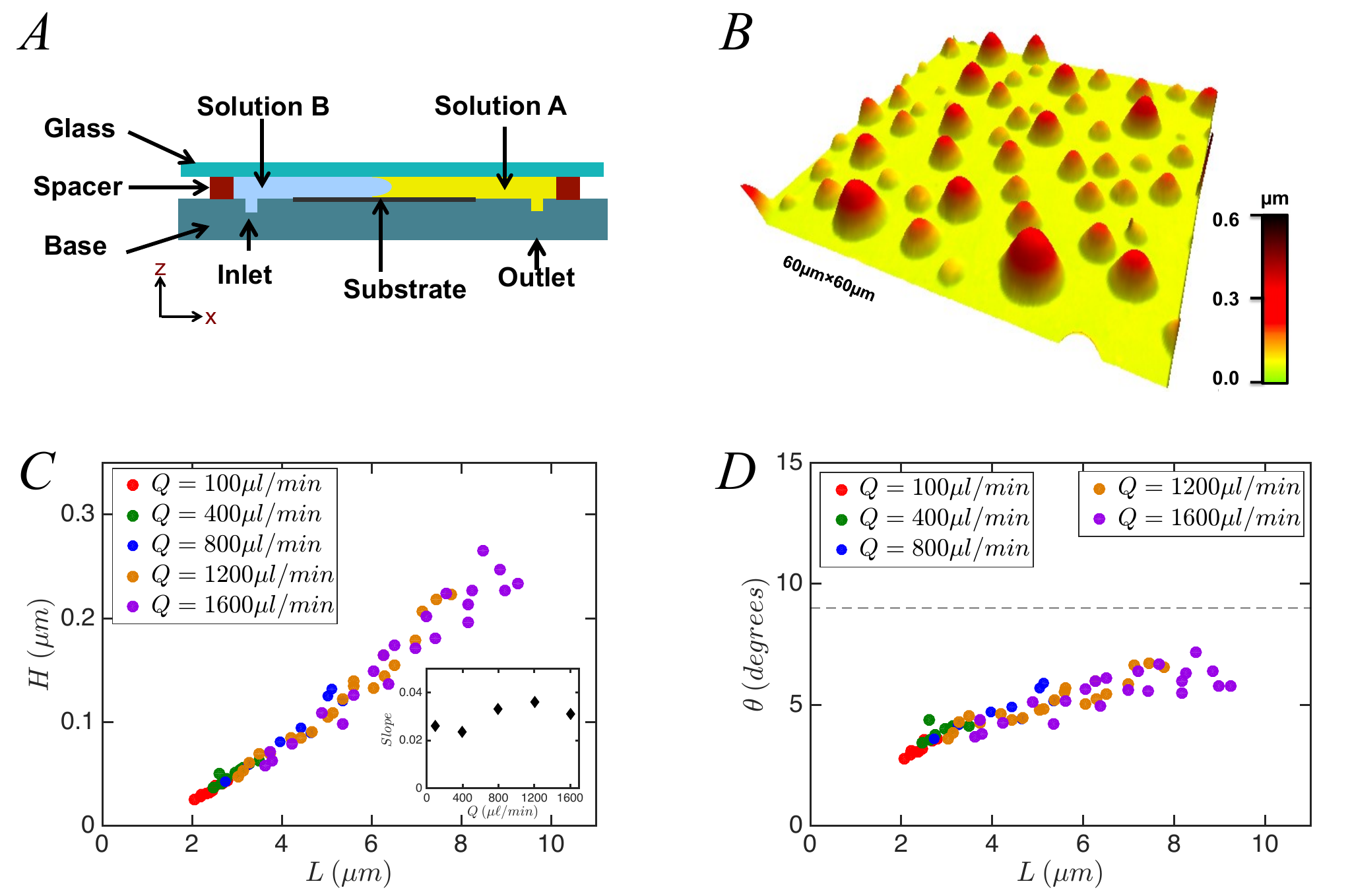}}
\caption{Solvent exchange process and morphologic features of surface droplets formed at different flow rates for the narrow 
channel ($h$ = 0.33 mm). {\it (a)} Schematic drawings of a fluid channel. The  channel consists of a glass top window, spacer and a base. The hydrophobic substrate is placed inside the cell, facing to the transparent glass window. The distance between the substrate and the 
glass bottom surface can be adjusted by the thickness of the spacer.  The flow direction is in x direction.  %Note that both solutions are subject to the no-slip flow boundary conditions on the flow walls and thus also on the substrate.
  {\it (b)} shows a representative AFM image of the polymerized droplets.
 {\it (c)} Droplet height, H, and 
{\it (d)} the contact angle, $\theta$, versus the lateral diameter, L, of the nanodroplets. The macroscopic advancing and receding angles are labelled for comparison. 
 }
\label{afm}
\end{figure*}

\begin{figure*}
\centerline{\includegraphics[width=.7\textwidth]{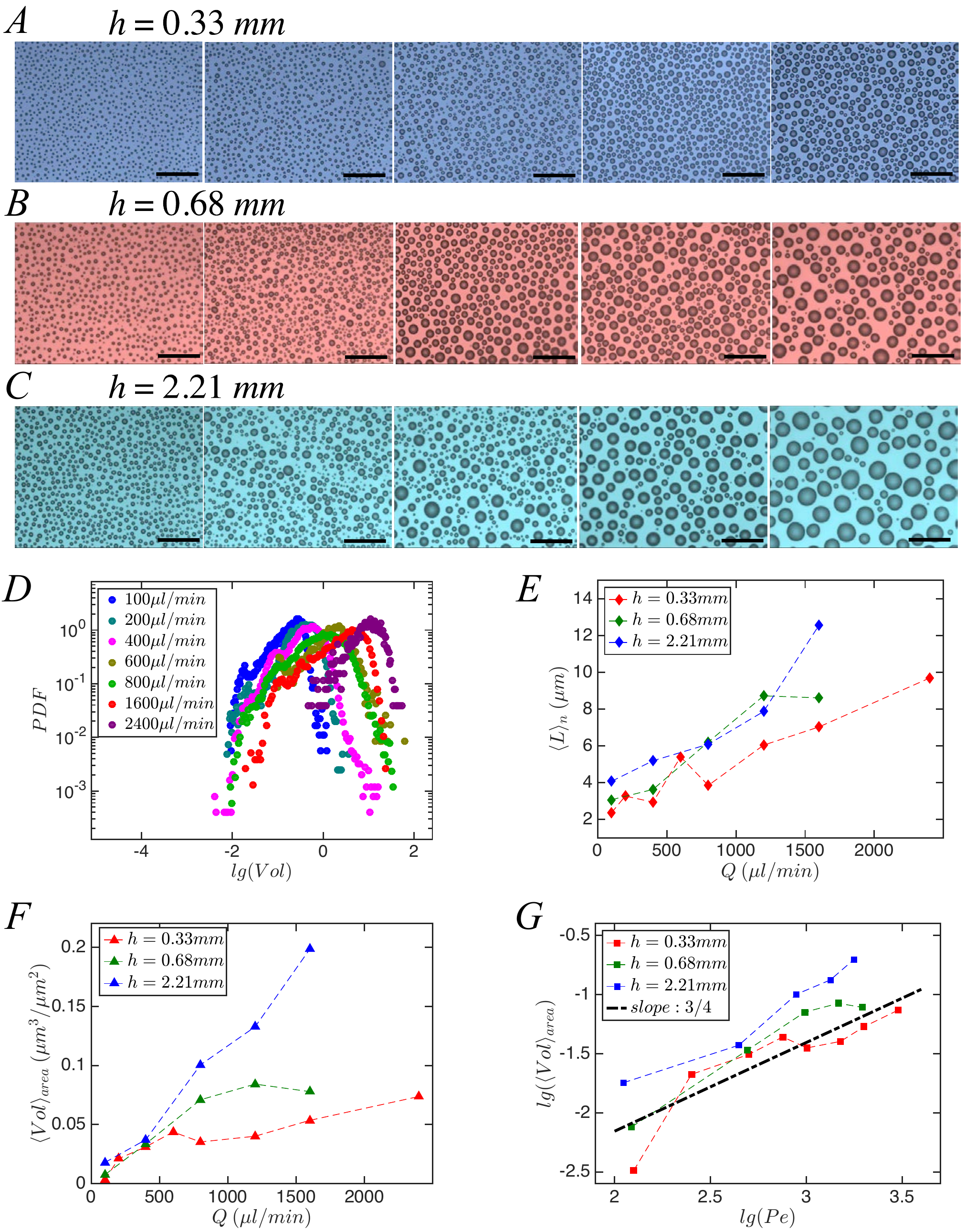}}
\caption{Optical images and size of surface droplets formed at different flow rates in three channels. 
{\it(a-c)} shows reflection-mode optical images of the polymerised droplets at different flow rates.
 Length of scale bars: 50 $\mu$m.  The flow rates were 100 $\mu$l/min, 400 $\mu$l/min, 800 $\mu$l/min, 1200 $\mu$l/min, and 1600 $\mu$l/min.
 The plots show the PDF of the droplet volume produced in the narrowest channel {\it (d)}, 
the averaged lateral diameter {\it(e)} and the averaged volume {\it (f)} of the surface 
 droplets at different flow rates. 
 %{\it (c)} shows the average droplet diameter $\langle L \rangle_{n}$ versus the channel height  $h$ at different flow rates.
 {\it (g)} shows  the averaged volume of droplets per $\mu$m$^{2}$  as function of the
Peclet number on a log-log plot. The dashed line shows
the scaling law $<Vol>_{area} \propto Pe^{3/4}$.}
\label{optical}
\end{figure*}

\begin{figure*}
\centerline{\includegraphics[width=.9\textwidth]{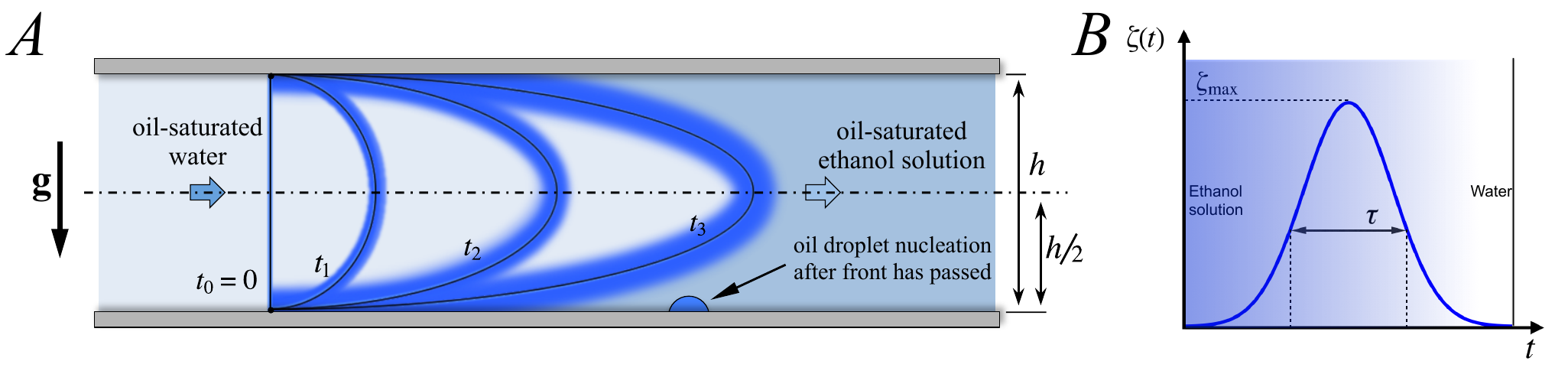}}
\caption{
	{\it (a)} Parabolic flow profiles for various times $t\ge 0$. Note the no-slip boundary condition for the flow.
	The originally sharp interface will broaden with time.  {\it (b)} Approximate temporal evolution of the oversaturation $\zeta (t)$ at fixed position downstream.
	The width $\tau$ of the pulse is defined through $\int_{-\infty}^\infty \zeta (t) dt = \zeta_{max} \tau$. % {\it (c)} Sketch and notation of a nucleated droplet.
	}
   \label{profile}
\end{figure*}

\begin{figure*}
\centerline{\includegraphics[width=.9\textwidth]{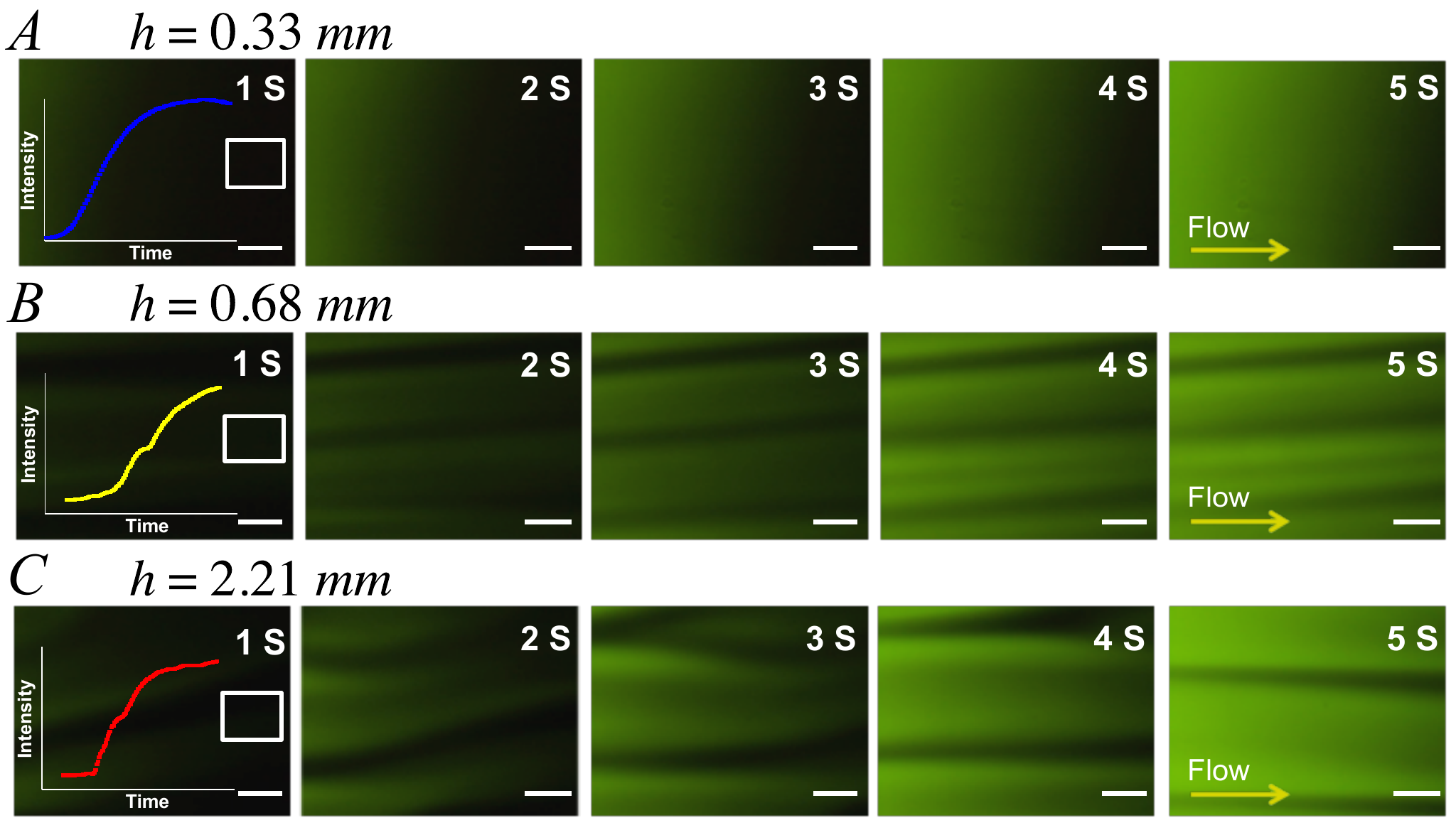}}
\caption{Top-view snapshots of the flow during the solvent exchange in three channel heights. The direction of the water (dyed green) in the fluorescent images was from left to right. Length of the scale bar: 200 $\mu$m. Insert is the curve of integrated optical density of selected area as function of time as water was pushed through the channel. The mean flow velocity $\bar U= 0.36$  mm/s in all three channels. } 
\label{snapshot}
\end{figure*}

\begin{figure*}
\centerline{\includegraphics[width=.9\textwidth]{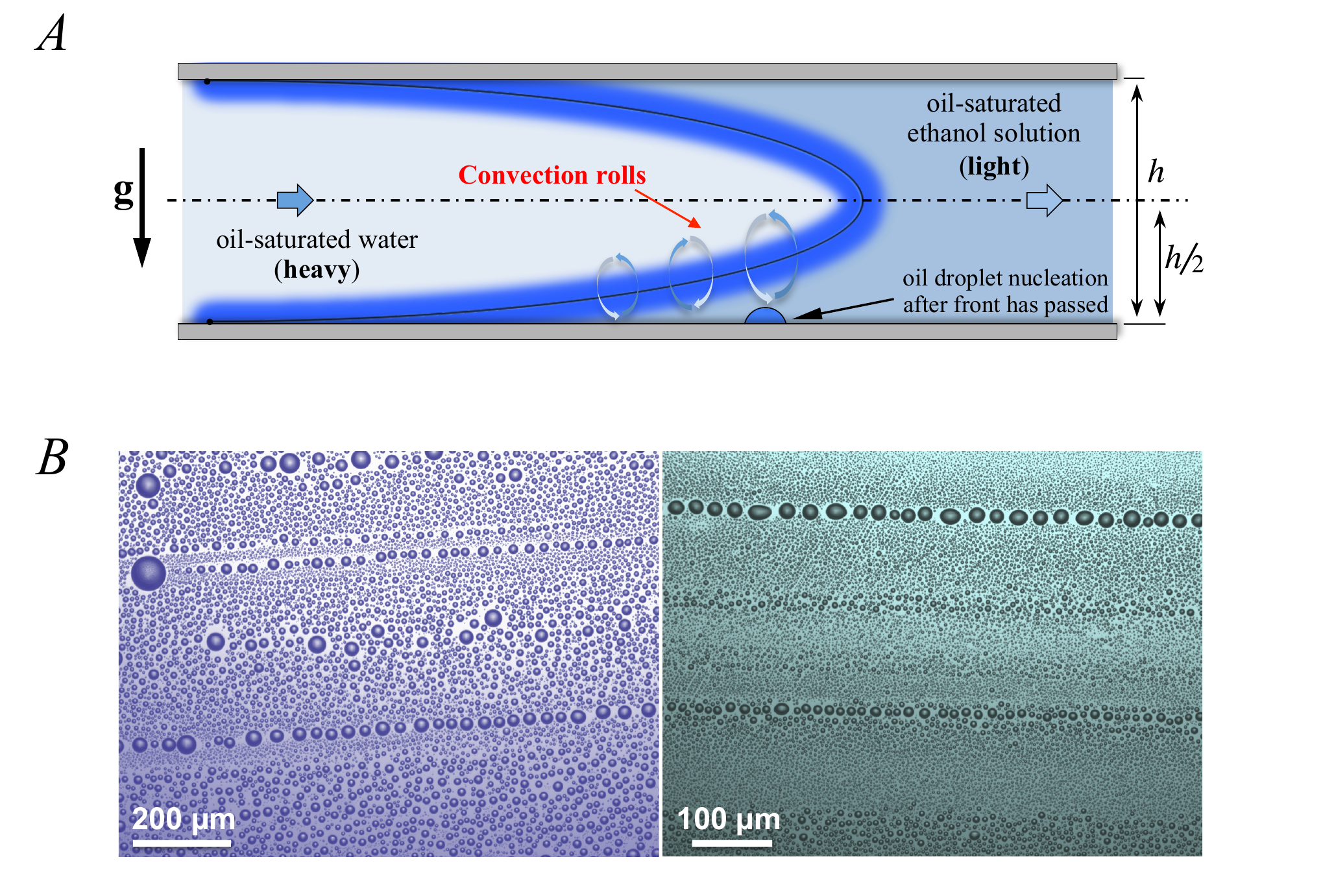}}
\caption{Solvent exchange and nanodroplet lines in the less narrow channels. % {\it (a)(b)} Top-view sequences of fluorescent video frames of tracking solvent exchange.  Length of the scale bar: 200 $\mu$m.  {\it (a)}: $h=0.68$ mm and $Q= 200 \mu$l/min.  {\it (b)}: $h=2.21$ mm and $Q= 670 \mu$l/min. The mean flow velocity $\bar U$ was 0.36 mm/s for both channels. The stripy structure reflects the existence of convections rolls with axis parallel to the flow direction.  Insert is the curve of integrated optical density of selected area with time. 
{\it (a)} The illustration shows convection rolls in the bottom part of interface between two solutions during the solvent exchange, where the heavy oil-saturated water is above the light oil-saturated ethanol aqueous solution.
{\it (b)} Two representative images of the nanodroplet lines formed in two less narrow channels. Left panel: $Q=1600 \mu$l/min and $h=0.68$ mm. Right panel: $Q=400 \mu$l/min and $h=2.21$ mm. The droplets organise in rows, reflecting the 
convection rolls with axes in flow direction. These rolls enhance mixing, leading to larger droplets where the rolls hit the 
substrate.}
\label{convection}
\end{figure*}

The experimental setup is shown in the schematic drawing Figure~\ref{afm}(a). A flow cell was constructed by assembling a glass plate, a spacer and a base together, forming a channel where the OTS-Si was put in. The channel height between the OTS-Si substrate and glass plate was adjusted by the thickness of the spacer. 5 ml solution A was first injected into the flow cell, followed by the injection of 10 ml solution B with a constant flow rate controlled by a syringe pump. After the formation of the nanodroplets, the flow cell was illuminated under an UV lamp (20 W, 365 nm) for 15 min, allowing the polymerization of the monomer droplets. The substrate was then washed with ethanol and dried by a gentle stream of nitrogen.  Images of the polymerised microdroplets were acquired using a reflection-mode optical microscopy. High resolution images of the polymerised microdroplets were also obtained from normal contact mode AFM imaging in air (Asylum Research, Santa Barbara, CA).

\begin{addendum}
 \item We thank Andrea Prosperetti for 
making us aware of the possible occurrence of convection rolls in this system and for 
very stimulating discussions, and Shuhua Peng for assisting in the experiments.  X.H.Z. acknowledges the support from Australian Research Council (FT120100473, DP140100805), L.B. from Endeavour Research Fellowship. D.L. gratefully acknowledges the support from an ERC-Advanced Grant and MCEC.  \item[Competing Interests] The authors declare that they have no
competing financial interests.
 \item[Correspondence Email]  d.lohse@utwente.nl.
\end{addendum}

\clearpage
\newpage

\begin{table*}[h]
\centering
\caption{Experimental cases and parameters for different fluid channel heights $h$. 
For all cases the width of fluid channels $w$ and the length $l$ are 14 mm and 56 mm,  respectively.
$Q$ is the flow rate, $\bar{U}=Q / (wh) $ the mean flow velocity, 
$Re=\bar U h/\nu = Q / (w\nu ) $ the Reynolds number, and 
$Pe=\bar U h/ D = Q / (w D ) $ the Peclet number. %Both $Pe$ and $Re$ are independent of the channel height $h$ for given flow rate.
}
\begin{tabular}{@{\vrule height 9pt depth2pt  width0pt}rccccc}
\vrule depth 6pt width 0pt {\it Q ($\mu$l/min)}& {\it Re}& {\it Pe}&\multicolumn3c{\it $\bar{U}$ (mm/s)}
 \\
\hline
\noalign{\vskip-13pt}
\\
& & & {\it h=0.33mm} & {\it h=0.68mm} & {\it h=2.21mm}\\

100& 0.04&119&0.36&0.18&0.05\\
200& 0.09&238&0.72&0.35&0.11\\
400& 0.18&476&1.44&0.70&0.22\\
600& 0.27&714&2.16&1.05&0.32\\
800& 0.35&952&2.88&1.40&0.43\\
1200& 0.53&1429&4.32&2.10&0.65\\
1600& 0.71&1905&5.77&2.80&0.86\\
2400& 1.06&2857&8.66&4.20&1.29\\
\hline
\end{tabular}
\label{tab:geometry}
\end{table*}

\end{document}